# How can AI reduce fall injuries in the workplace?

N. Cartocci[1,2], A. E. Gkikakis[1], R.F. Pitzalis[1,3], F. Pera[4], M.T. Settino[4], D. G. Caldwell[1], J. Ortiz[1]
[1]XoLab, Advanced Robotics, Istituto Italiano di Tecnologia (IIT), Genoa, Italy
[2]Department of Informatics, Bioengineering, Robotics and Systems Engineering (DIBRIS), University of Genoa (UniGe), Genoa, Italy
[3]Department of Mechanical, Energy and Transportation Engineering (DIME), University of Genoa, 16145, Genoa, Italy.
[4]Department of Technological Innovations and Safety of Plants, Products, and Anthropic Settlements, INAIL, Rome, Italy

Speaker: N. Cartocci

## Abstract

Fall-caused injuries are common in all types of work environments, including offices. They are the main cause of absences longer than three days, especially for small and medium-sized businesses (SMEs). However, data, data amount, data heterogeneity, and stringent processing time constraints continue to pose challenges to real-time fall detection. This work proposes a new approach based on a recurrent neural network (RNN) for Fall Detection and a Kolmogorov-Arnold Network (KAN) to estimate the time of impact of the fall. The approach is tested on SisFall, a dataset consisting of 2706 Activities of Daily Living (ADLs) and 1798 falls recorded by three sensors. The results show that the proposed approach achieves an average TPR of 82.6% and TNR of 98.4% for fall sequences and 94.4% in ADL. Besides, the Root Mean Squared Error of the estimated time of impact is approximately $160 ms$.

## 1. Introduction

Every year, 684,000 people lose their lives, as reported by the World Health Organization (WHO). Deaths from falls have increased much faster than any other type of injury in the past two decades. In addition to the number of deaths, 172 million people are disabled by falls each year [1]. This increase is due to many factors, such as aging and urbanization patterns. Although fall-caused injuries are often overlooked, they have a variety of consequences not only for the fall victim and those within their surroundings but also for the entire health/economic system. Specifically, the US spent a total of 50 billion dollars on medical expenses due to falls among senior citizens [2]. Slip and fall injuries are common in all types of work environments, including offices, and are the main cause of absences longer than three days, especially for small and medium-sized businesses (SMEs) [3]. In Italy, falls at the same level cause serious injuries in workers with an average absence duration of 38 days, a duration exceeded only by those due to falls from heights of 47 days. The loss of approximately 2.5 million working days in all sectors is one of the main causes of unemployment, which has a serious negative economic impact on the entire national production system. Compensation for fall injuries is more than 90 million euros (direct costs) and is one of the primary expenses of the Italian Workers' Compensation Authority (INAIL). Since, as a first approximation, indirect costs can be approximately three times the direct costs, the total cost of fall injuries is more than 370 million euros per year [4]. A key aspect of developing a fall prevention system is to detect and/or predict falls. These technologies can now be found in wearable devices thanks to recent sensors and integrated circuit advances. However, data quality (in the presence of interference and unstructured surroundings), data amount, data heterogeneity, and stringent processing time constraints (fall must be identified within a tight time restriction) continue to pose challenges to real-time fall detection [5]. Most studies focus on the detection of falls after impact [6], [7], [8], [9]; instead, this research aims to find a real-time software solution that can be integrated into a plug-and-play wearable device to mitigate impact. The software solution should be able to adjust to the needs of various users, irrespective of their age, gender, or physical makeup. It should also function well in most environments, protecting workers from injury if they fall to the same or lower level. Unlike previous studies [10], [11], [12], [13], which focus on older people, this work aims to address a more agile group of people, industrial workers. Workers on construction sites are more at risk and are forced to perform repetitive, heavy, and challenging tasks, which have a greater impact on their mental and physical health than in other industries. In addition, most studies ignore the subject's reaction. For example, young and healthy subjects will not fall like older people, but in many cases, they may be fast enough to arrest the fall or catch objects around them. This introduces false positive effects, causing improper activation of the prevention system, and may discourage users from wearing it. The appropriate activation of the prevention mechanism is the main goal; fall detection alone is insufficient for our study. Therefore, this work considers the critical issues and opportunities highlighted above and proposes a new approach based on a recurrent neural network (RNN) for Fall Detection and a Kolmogorov-Arnold Network (KAN) to estimate the time of impact of the fall [14]. Effectively predicting the time of impact (i.e., lead time) during a fall is not yet adequately addressed. Some approaches rely on predicting the trajectory of the fall to estimate the time of impact [15], [16], [17].

## 2. SisFall

The publicly available SisFall [18] dataset is used to train the proposed approach because it is one of the most comprehensive datasets with many subjects and falls. It contains 15 types of falls and 19 types of Activities of Daily Living (ADLs), listed in Tables 1 and 2.

| Code | Activity |
|---|---|
| F01 | Fall forward while walking caused by a slip |
| F02 | Fall backward while walking caused by a slip |
| F03 | Lateral fall while walking caused by a slip |
| F04 | Fall forward while walking caused by a trip |
| F05 | Fall forward while jogging caused by a trip |
| F06 | Vertical fall while walking caused by fainting |
| F07 | Fall while walking, with the use of hands-on a table to dampen fall, caused by fainting |
| F08 | Fall forward when trying to get up |
| F09 | Lateral fall when trying to get up |
| F10 | Fall forward when trying to sit down |
| F11 | Fall backward when trying to sit down |
| F12 | Lateral fall when trying to sit down |
| F13 | Fall forward while sitting, caused by fainting or falling asleep |
| F14 | Fall backward while sitting, caused by fainting or falling asleep |
| F15 | Lateral fall while sitting, caused by fainting or falling asleep |

*Table 1 - Types of falls selected in SisFall.*

| Code | Activity |
|---|---|
| D01 | Walking slowly |
| D02 | Walking quickly |
| D03 | Jogging slowly |
| D04 | Jogging quickly |
| D05 | Walking upstairs and downstairs slowly |
| D06 | Walking upstairs and downstairs quickly |
| D07 | Slowly sit in a half-height chair, wait a moment, and up slowly |
| D08 | Quickly sit in a half-height chair, wait a moment, and up quickly |
| D09 | Slowly sit in a low-height chair, wait a moment, and up slowly |
| D10 | Quickly sit in a low-height chair, wait a moment, and up quickly |
| D11 | Sitting a moment, trying to get up, and collapse into a chair |
| D12 | Sitting a moment, lying slowly, wait a moment, and sit again |
| D13 | Sitting a moment, lying quickly, wait a moment, and sit again |
| D14 | Being supine, moving to a lateral position, waiting a moment, and moving back |
| D15 | Standing, slowly bending at knees, and getting up |
| D16 | Standing, slowly bending without bending knees, and getting up |
| D17 | Standing, get into a car, remain seated, and get out of the car |
| D18 | Stumble while walking |
| D19 | Gently jump without falling (trying to reach a high object) |

*Table 2 - Types of activities of daily living selected in SisFall.*

The activities were recorded by 38 volunteers, 23 adults, and 15 elderly (60-75 years old). The elderly simulated only ADLs, except a Judo expert, who also simulated falls. In addition, some elderly did not do some activities due to personal problems (or medical recommendations). Each task was repeated 5 times for each subject. Overall, the dataset includes 2706 ADLs and 1798 falls recorded by three sensors (two accelerometers: an Analog Devices ADXL345 and a Freescale MMA8451Q, and one gyroscope Texas Instruments ITG3200) worn on the waist at $200 Hz$. Accurate labeling of the fall period from the motion sequence is a time-consuming and labor-intensive prerequisite for algorithm development. To annotate the fall period from the SisFall dataset, Musci et al. [19] first defined the standard labeling criteria by consulting experts in the field, then performed the labeling separately and revised it jointly in multiple steps to improve consistency. This process tends to be subjective because SisFall was constructed without video references.

## 3. Fall Detection Data Processing

A deep recurrent neural network (RNN) was chosen for data-based fall detection, where the data used to train the neural network are obtained from the SisFall dataset [18] labeled in [19] as no-background. Using the two accelerometers and one gyroscope, the orientation of the subject was calculated through a six-axis Kalman filter [20] and expressed as a quaternion. Then, the angle $\theta$ and acceleration $\dot{\theta}$ of the body relative to the ground were estimated. A full description of the features used is given in Table 3.

| Symbol | Description |
|---|---|
| $y$ | Age of the subject |
| $h$ | Height of the subject |
| $w$ | Weight of the subject |
| $g$ | Gender of the subject |
| $a_{x,ADXL345}$ | Acceleration data in the X axis measured by the sensor ADXL345 |
| $a_{y,ADXL345}$ | Acceleration data in the Y axis measured by the sensor ADXL345 |
| $a_{z,ADXL345}$ | Acceleration data in the Z axis measured by the sensor ADXL345 |
| $a_{x,MMA8451Q}$ | Acceleration data in the X axis measured by the sensor MMA8451Q |
| $a_{y,MMA8451Q}$ | Acceleration data in the Y axis measured by the sensor MMA8451Q |
| $a_{z,MMA8451Q}$ | Acceleration data in the Z axis measured by the sensor MMA8451Q |
| $\dot{\alpha}_{x,ITG3200}$ | Rotation data in the X axis measured by the sensor ITG3200 |
| $\dot{\alpha}_{y,ITG3200}$ | Rotation data in the Y axis measured by the sensor ITG3200 |
| $\dot{\alpha}_{z,ITG3200}$ | Rotation data in the Z axis measured by the sensor ITG3200 |
| $q_1$ | Orientation data in Quaternion, the first element |
| $q_2$ | Orientation data in Quaternion, the second element |
| $q_3$ | Orientation data in Quaternion, the third element |
| $q_4$ | Orientation data in Quaternion, the fourth element |
| $\theta$ | Body angle relative to the direction of gravity |
| $\dot{\theta}$ | Body angular acceleration relative to the direction of gravity |

*Table 3 - List of features*

The network has multiple inputs, consisting of 4 inherent characteristics of the subjects: Age, Height, Weight, Gender, and a multivariate time series consisting of the measurements of the two accelerometers and the gyroscope, the orientation of the body (expressed as a quaternion), and the relative angle of the body to the direction of gravity $\theta$. In total, the neural network input has 18 entries. The output is the probability of falling $P(falling)$, and if $P(falling) > 0.5$, the subject is assumed to be falling. The network architecture is chosen similarly to the RNN architecture used in [19] and depicted in Figure 1. The input is processed by the fully connected Layer 1, and a second fully connected layer (Layer 8) collects the output from the LSTM cell of Layer 6. Its output is sent to the final SoftMax classifier (Layer 9), which produces the falling probability of the classification. The core of the network is based on the two LSTM cells stacked in layers (Layers 4 and 6). LSTM cells are popular Deep Learning architectures because of their ability to capture time-dependent relationships in the input data efficiently. The size of the tensors within each LSTM gate, commonly referred to as the inner dimension, is chosen to be $2^4$, as well as the neurons in the first fully connected layer. The architecture also includes a batch normalization layer (Layer 2) to regularize the input data and three drop-out layers (Layers 3, 5, and 7). The latter layers are used during training to improve generalization as they are removed from the inference module built into the device. The dropout rate is selected to be 50%.

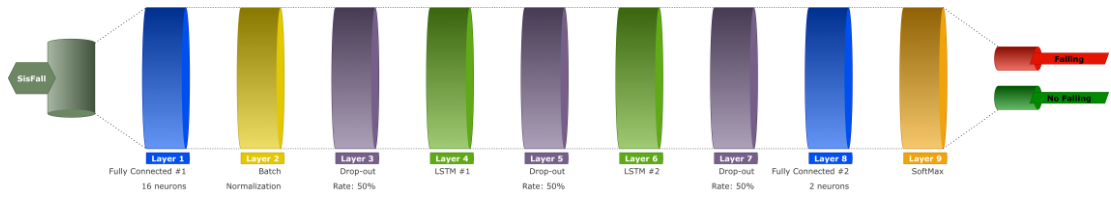

*Figure 1 - The architecture of the Fall Detection Neural Network.*

The portion of the dataset used to train, validate, and test the FDNN contains only the sequences with a fall, 1798 fall sequences for 24 subjects, 23 adults, and one elderly. Then, this standardized dataset is shuffled and divided into 3 sub-datasets, one for training, one for validation, and one for testing the FDNN algorithm, in 60%, 20%, and 20%, respectively. The FDNN is trained using the training dataset to train the weight of the FDNN and the validation dataset to choose the optimal weights, that is when the accuracy of the FDNN on the Validation dataset is highest. In the training procedure, the batch size is set to 128 sequences and the number of epochs is set to 64.

## 4. Time of impact Data Processing

Overall, 19 signals are considered at the beginning of this study. From each 15-second-long experiment, only the part related to the fall was extracted, thus excluding the initial activity, e.g., walking, jogging, sitting, etc., and the final parts where the subject remains motionless after hitting the ground. This interval was defined from the time the data are labeled by Musci et al. [19] as "FALL" to the time when the standard deviation of the acceleration is less than a threshold. The time of impact was defined as the time between the moment under investigation and the moment when the subject hits the ground. This value is a multiple of the sampling period, i.e., $5ms$, and decreases during the fall until it reaches zero.

### 4.1 Feature Selection

A fully data-driven approach is adopted based on the use of two feature selection algorithms to restrict the number of features. Specifically, using correlation analysis and a feature selection algorithm called Minimum Redundancy Maximum Relevance (mRMR) [21]. Correlation analysis is the simplest and, very often, one of the most effective methods for feature reduction. It can be used in a variety of ways, such as when two features have a high correlation, one of them can be eliminated, or if one feature is poorly correlated with the value you want to estimate, that feature can be eliminated. In this study, the second approach was considered and the features below a correlation score were excluded. The mRMR algorithm identifies a set of mutually dissimilar and maximally distinct features, allowing them to effectively represent the response variable. The algorithm seeks to maximize the relevance of the features to the response variable while minimizing redundancy within the feature set. The method makes use of mutual information measures, such as mutual information between a feature and the response variable, as well as mutual information of the variables and features. Through the correlation analysis, four features were found to be the most highly correlated with the time of impact and $a_{y,ADXL345}$,

$a_{y,MMA8451Q}$, $\theta$ and $\dot{\theta}$ were chosen. Similarly, using the mRMR algorithm, the selected features are $\dot{\alpha}_{y,ITG3200}$ and $\theta$.

## 4.2 Fine Tuning

The KAN identification problem has been solved by using piecewise linear basis functions as supporting basis functions and the Newton-Kaczmarz method [22] for solving systems of nonlinear equations. However, it is necessary to decide other parameters of the model, such as the number of internal and external piecewise linear basis functions, $n$ and $q$, respectively, the learning rate or regularization parameter $\mu$ of the Newton-Kaczmarz method, and the time window $w$ used to predict the time of impact. All these parameters are calculated through a k-fold cross-validation process. Each subject repeated the task 5 times, so three repetitions are used to train the model and one to validate it. The last repetition was used to test the performance of the proposed approach and compare it with other well-known Machine Learning methods. This approach was used to reduce data overfitting and train with as much varied data as possible. The training processing lasted 10 epochs, and the data were standardized by subtracting from the mean and dividing by standard deviation to put different variables on the same scale. The best combination of parameters was found to be $n = 4$; $q = 64$; $\mu = 0.0625$; and $w = 50\ ms$.

# 5. Results

This section describes the results obtained considering our proposed RNN for Fall Detection and KAN for Time of Impact Estimation. The results below relate to the data in SisFall and analyze the performance of the approach both with respect to different activities and for different subjects.

# 6. Fall Detection

Once the FDNN is trained, it is tested on the Test dataset (~360 sequences). Figures 2 and 3 show the TNR and TPR as a function of the subjects and the activities, respectively. The tables show the average value between the repetitions, and the blank cells mean that there are no samples of those subjects and activities in the Test dataset. In particular, the subjects SE01-SE05 do not simulate the Falling events, as written in the SisFall paper description. Being the annotation a manual procedure, the start and end of the actual fall may not precisely match the time defined in the annotated dataset. The TNR value is, in most cases, more than 95% and, on average, 98.4%. The TPR value is, on average, 82.6%. In many cases, the FDNN succeeds in even perfectly classifying the fall; however, there are some cases in which the TPR is not much higher than 50%. Besides, it is important to consider that since the fall is an event that persists for a very short time, failure to classify even a few samples lowers the TPR significantly. Analyzing the different activities, activity F10 appears to be the most difficult to classify correctly. Then, the FDNN is tested on the sequences of the SisFall dataset without fall (ADLs), 2701 sequences. Figure 4 shows the TNR and TPR, respectively, as a function of subject and activity. The TNR value is, in most cases, 100%; but in some cases, it is also less than 20%. Particularly for activities: D03, D04, D12, D13 and D14. The worst activity for the TNR metric is D04. The average TNR value is 94.4%.

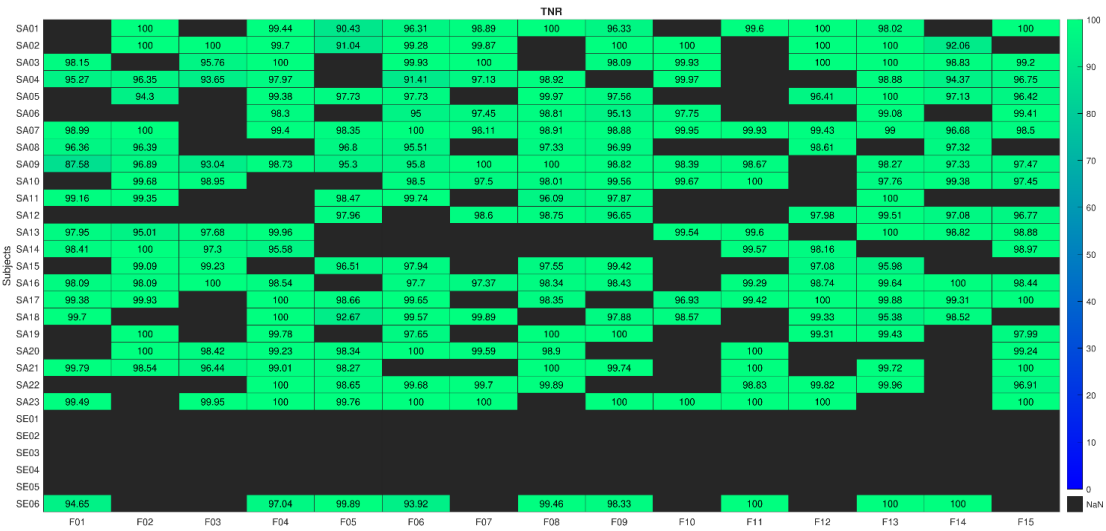

*Figure 2 - TNR as a function of subject and activity for falls.*

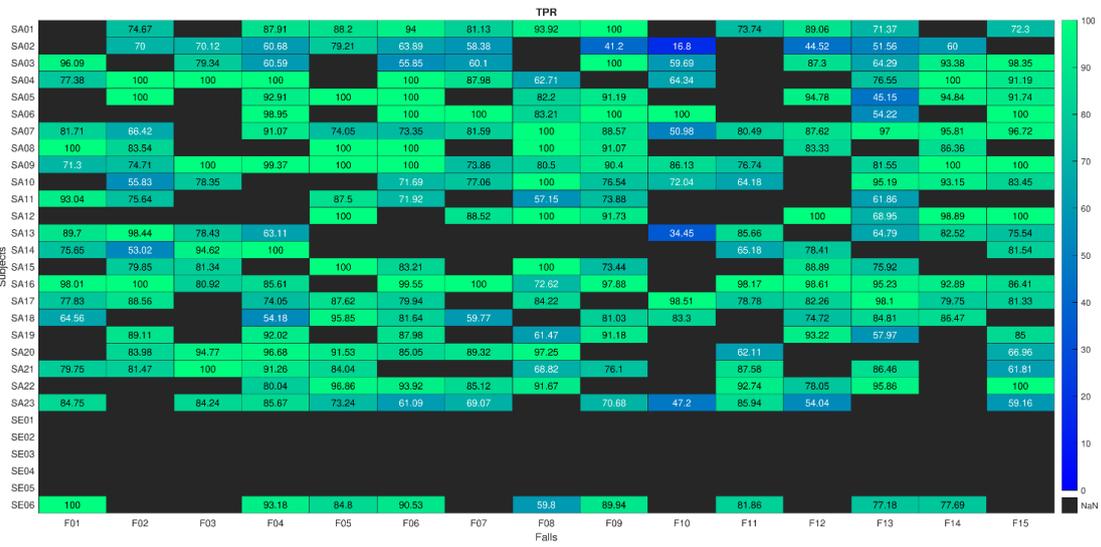

*Figure 3 - TPR as a function of subject and activity for falls.*

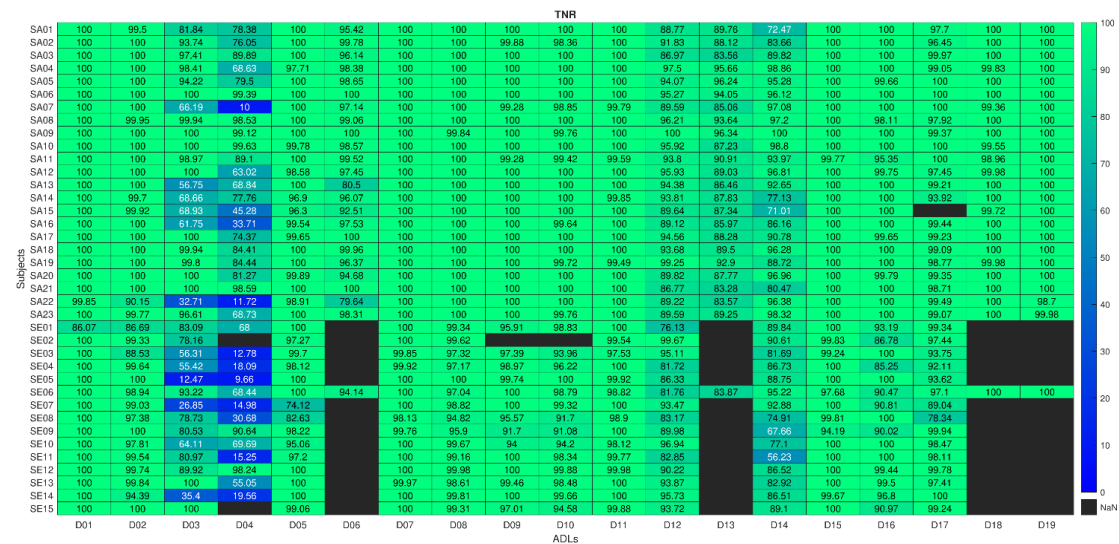

*Figure 4 - TNR as a function of subject and activity for ADLs.*

## 7. Time of Impact Estimation

Figure 5 shows the RMSE for different subjects and falls, considering the best parameters on the test data. The performance obtained is generally positive, on average the RMSE on the 1798 falls is about 160 $ms$. On some subjects, the algorithm obtains worse results, such as for SA10; likewise, some types of falls, such as F13. Particularly in the combinations SA05-F13 and SA21-F11, the RMSE exceeds the mean error by almost 4 times. In contrast on F05 and F15, errors are never found to be too large, demonstrating that the proposed KAN is an excellent method for estimating time of impact and suitable for estimation in the case of falls of different types. With SA03, SA06, and SA12, the proposed approach achieves excellent results for almost all types of falls.

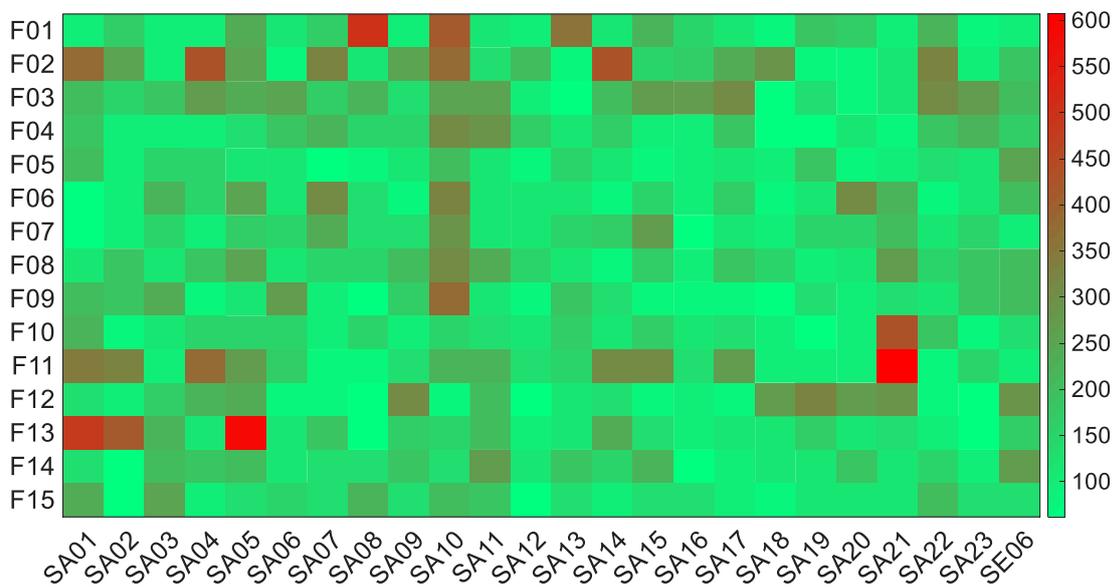

*Figure 5 - Heatmap about RMSE for different subjects and falls [ms].*

Finally, the performance obtained by the proposed KAN is evaluated, showing the trend of true time of impact (Ground Truth) and projected time of impact. In each instant, the predicted time of impact is estimated. Figure 6 shows this trend in the case of the 18th subject, who lateral falls while walking caused by a slip (F3). The fall takes about 700 $ms$, and the ground truth is a unit line from this value to 0, which is the moment the subject touches the ground. When the values of time of impact are sufficiently large, the network faithfully follows the trend. On the other hand, when the time of impact is very small, the network estimates a constant value of approximately 140 $ms$.

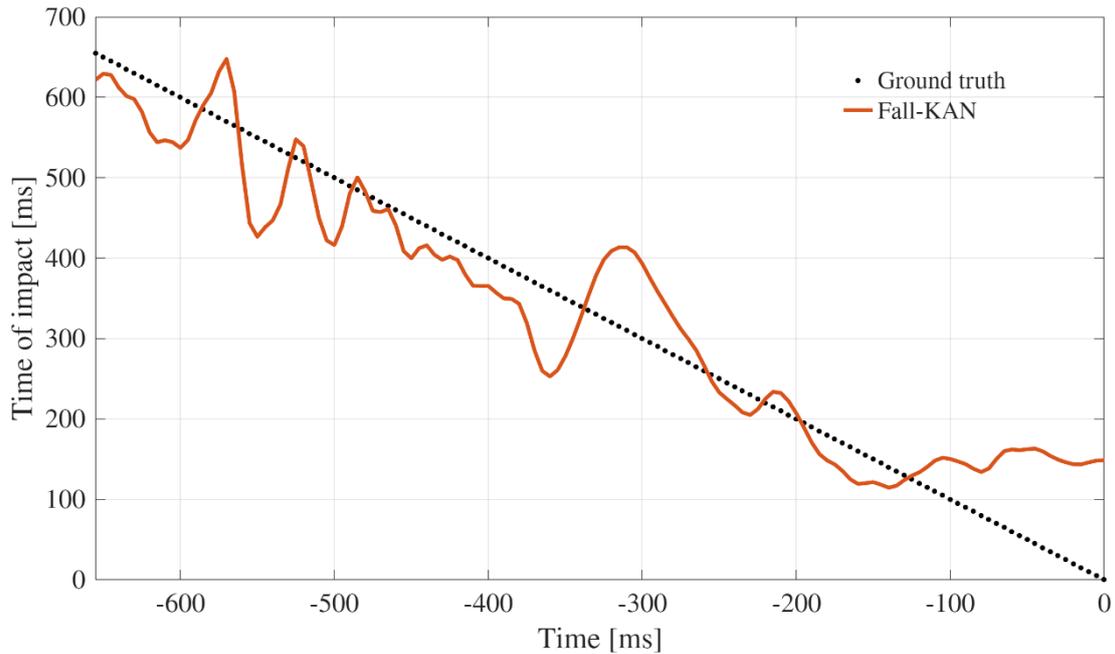

*Figure 6 - Time of impact for the 18th subject during the 3rd fall.*

## 8. Conclusions

The Fall Detection Neural Network performed excellently on SisFall with an average TPR of 82.6% and TNR of 98.4% in fall sequences and 94.4% in ADL sequences. While with the KAN, only five features were found to be useful for time of impact estimation, of which 3 were already in the dataset, and two were calculated using an accelerometer and a gyroscope. The RMSE obtained by the proposed KAN is about $160 ms$. However, this value is distorted by estimating very small values of impact time where the error is constant and equal to about $140\ ms$ in the example presented. Future studies will focus on the joint use of the two techniques and performance evaluation.

## 9. References

[1] J. D. Stanaway *et al.*, "Global, regional, and national comparative risk assessment of 84 behavioural, environmental and occupational, and metabolic risks or clusters of risks for 195 countries and territories, 1990–2017: a systematic analysis for the Global Burden of Disease Study 2017," *The Lancet*, vol. 392, no. 10159, pp. 1923–1994, Nov. 2018, doi: 10.1016/S0140-6736(18)32225-6.

[2] C. S. Florence, G. Bergen, A. Atherly, E. Burns, J. Stevens, and C. Drake, "Medical Costs of Fatal and Nonfatal Falls in Older Adults," *J Am Geriatr Soc*, vol. 66, no. 4, pp. 693–698, Apr. 2018, doi: 10.1111/jgs.15304.

[3] "Step safely: strategies for preventing and managing falls across the life-course. Geneva: World Health Organization," 2021.

[4] A. CURA *et al.*, "La valutazione ergonomica del rischio caduta in piano".

[5] N. Cartocci *et al.*, "Recognition of Physiological Patterns during Activities of Daily Living Using Wearable Biosignal Sensors," in *22nd Congress of the International Ergonomics Association (IEA 2024)*, Jeju: Springer, Aug. 2024.

[6] J. Pena Queralta, T. N. Gia, H. Tenhunen, and T. Westerlund, "Edge-AI in LoRa-based health monitoring: Fall detection system with fog computing and LSTM recurrent neural networks," *2019*

[ ]  *42nd International Conference on Telecommunications and Signal Processing, TSP 2019*, pp. 601–604, Jul. 2019, doi: 10.1109/TSP.2019.8768883.
[7]   A. Chelli and M. Patzold, "A Machine Learning Approach for Fall Detection and Daily Living Activity Recognition," *IEEE Access*, vol. 7, pp. 38670–38687, 2019, doi: 10.1109/ACCESS.2019.2906693.
[8]   A. Ramachandran and A. Karuppiah, "A Survey on Recent Advances in Wearable Fall Detection Systems," *Biomed Res Int*, vol. 2020, 2020, doi: 10.1155/2020/2167160.
[9]   S. Usmani, A. Saboor, M. Haris, M. A. Khan, and H. Park, "Latest research trends in fall detection and prevention using machine learning: A systematic review," *Sensors*, vol. 21, no. 15, p. 5134, Jul. 2021, doi: 10.3390/s21155134.
[10]  G. Wang, Q. Li, L. Wang, Y. Zhang, and Z. Liu, "Elderly fall detection with an accelerometer using lightweight neural networks," *Electronics (Switzerland)*, vol. 8, no. 11, p. 1354, Nov. 2019, doi: 10.3390/electronics8111354.
[11]  M. Saleh and R. L. B. Jeannes, "Elderly Fall Detection Using Wearable Sensors: A Low Cost Highly Accurate Algorithm," *IEEE Sens J*, vol. 19, no. 8, pp. 3156–3164, Apr. 2019, doi: 10.1109/JSEN.2019.2891128.
[12]  T. Namba and Y. Yamada, "Fall Risk Reduction for the Elderly by Using Mobile Robots Based on Deep Reinforcement Learning," *Journal of Robotics, Networking and Artificial Life*, vol. 4, no. 4, p. 265, 2018, doi: 10.2991/jrnal.2018.4.4.2.
[13]  C. J. Lord and D. P. Colvin, "Falls in the elderly: Detection and assessment," *Proceedings of the Annual Conference on Engineering in Medicine and Biology*, vol. 13, no. pt 4, pp. 1938–1939, 1991, doi: 10.1109/iembs.1991.684830.
[14]  N. Cartocci, A. E. Gkikakis, F. Pera, M. T. Settino, D. G. Caldwell, and J. Ortiz, "Fall-KAN: Fall impact time estimation Kolmogorov-Arnold Network," in *2024 4th International Conference on Electrical, Computer, Communications and Mechatronics Engineering (ICECCME)*, IEEE, Nov. 2024, pp. 1–6. doi: 10.1109/ICECCME62383.2024.10796549.
[15]  N. Cartocci, A. E. Gkikakis, D. G. Caldwell, and J. Ortiz, "Deep Learning-based wearable device to prevent fall from height injuries," Zenodo, Mar. 2024. doi: 10.5281/zenodo.10722452.
[16]  N. Cartocci, A. Gkikakis, D. Caldwell, and J. Ortiz, "Artificial intelligence-based wearable solution to prevent fall from heights injuries for the next generation of workers," in *2023 Slips, Trips and Falls (STF) International Conference*, Toronto, Jun. 2023.
[17]  N. Cartocci, A. Gkikakis, D. Caldwell, and J. Ortiz, "Real-time Fall Prevention system for the Next-generation of Workers," in *Workshop on Assistive Robotic Systems for Human Balancing and Walking: Emerging Trends and Perspectives @IROS 2022*, Kyoto, Oct. 2022.
[18]  A. Sucerquia, J. D. López, and J. F. Vargas-Bonilla, "SisFall: A fall and movement dataset," *Sensors (Switzerland)*, vol. 17, no. 1, p. 198, Jan. 2017, doi: 10.3390/s17010198.
[19]  M. Musci, D. De Martini, N. Blago, T. Facchinetti, and M. Piastra, "Online Fall Detection Using Recurrent Neural Networks on Smart Wearable Devices," *IEEE Trans Emerg Top Comput*, vol. 9, no. 3, pp. 1276–1289, Jul. 2021, doi: 10.1109/TETC.2020.3027454.
[20]  D. Roetenberg, H. J. Luinge, C. T. M. Baten, and P. H. Veltink, "Compensation of magnetic disturbances improves inertial and magnetic sensing of human body segment orientation," *IEEE Transactions on Neural Systems and Rehabilitation Engineering*, vol. 13, no. 3, pp. 395–405, Sep. 2005, doi: 10.1109/TNSRE.2005.847353.
[21]  Hanchuan Peng, Fuhui Long, and C. Ding, "Feature selection based on mutual information criteria of max-dependency, max-relevance, and min-redundancy," *IEEE Trans Pattern Anal Mach Intell*, vol. 27, no. 8, pp. 1226–1238, Aug. 2005, doi: 10.1109/TPAMI.2005.159.
[22]  M. Poluektov and A. Polar, "Construction of the Kolmogorov-Arnold representation using the Newton-Kaczmarz method," May 2023, [Online]. Available: http://arxiv.org/abs/2305.08194